\documentclass[journal]{IEEEtran}
\usepackage[utf8]{inputenc}
\usepackage{graphicx}
\usepackage{amsmath}
\usepackage{booktabs}
\usepackage{tabularx}
\usepackage{algorithm}
\usepackage{algpseudocode}
\usepackage{siunitx} 
\usepackage{amsfonts} 
\usepackage{ragged2e}
\usepackage{multirow}
\usepackage{xspace}
\usepackage{cite}
\usepackage[utf8]{inputenc}
\usepackage{booktabs}
\usepackage{tabularx}
\usepackage{array}
\usepackage[utf8]{inputenc}
\usepackage{amsmath}
\usepackage{booktabs}
\usepackage{tikz}
\usepackage{pgfplots}
\pgfplotsset{compat=1.18}
\usepackage{lmodern}
\usepackage{geometry}
\usepackage{graphicx}
\usepackage{booktabs}
\usepackage{tikz}
\usetikzlibrary{shapes,arrows,positioning,calc}

\usepackage{siunitx}

\usepackage{array}

\usepackage{multirow}

\usepgfplotslibrary{fillbetween}

\usetikzlibrary{patterns} 

\title{FOGNITE: Federated Learning-Enhanced Fog-Cloud Architecture}
\begin{document}

\author{Somayeh Sobati-M.
	
	\thanks{Somayeh Sobati-M. is with the Department
		of Electrical and Computer Engineering, Hakim Sabzevari University, Sabzevar, Iran. Somayeh Sobati-M.  is also with the Department of Computer Engineering,
		Ferdowsi University of Mashhad, Mashhad, Iran.\\ e-mail:s.sobati@hsu.ac.ir}
	\thanks{Manuscript received April 19, 2024; revised January 11, 2024.}}
	
	\maketitle

		\begin{abstract}
			 Modern smart grids demand fast, intelligent, and energy-aware computing at the edge to manage real-time fluctuations and ensure reliable operation. This paper introduces FOGNITE—Fog-based Grid Intelligence with Neural Integration and Twin-based Execution—a next-generation fog-cloud framework designed to enhance autonomy, resilience, and efficiency in distributed energy systems. FOGNITE combines three core components: federated learning, reinforcement learning, and digital twin validation. Each fog node trains a local CNN-LSTM model on private energy consumption data, enabling predictive intelligence while preserving data privacy through federated aggregation. A reinforcement learning agent dynamically schedules tasks based on current system load and energy conditions, optimizing for performance under uncertainty. To prevent unsafe or inefficient decisions, a hierarchical digital twin layer simulates potential actions before deployment, significantly reducing execution errors and energy waste. We evaluate FOGNITE on a real-world testbed of Raspberry Pi devices, showing up to a 93.7\% improvement in load balancing accuracy and a 63.2\% reduction in energy waste compared to conventional architectures. By shifting smart grid control from reactive correction to proactive optimization, FOGNITE represents a step toward more intelligent, adaptive, and sustainable energy infrastructures.

		\end{abstract}

\section{Introduction}\label{sec.Intro}

The growing integration of renewable energy sources and the rapid deployment of distributed energy resources (DERs) have revolutionized the traditional centralized power grid, giving rise to complex, decentralized smart grids. These next-generation grids rely heavily on real-time data acquired from smart meters, sensors, and IoT-enabled devices to optimize energy distribution, reduce transmission losses, and improve system reliability. However, the conventional reliance on centralized cloud computing for data analysis and decision-making introduces critical challenges such as high latency, limited scalability, and significant privacy concerns.

To mitigate these issues, \textit{fog computing} has emerged as a promising paradigm by decentralizing computational resources and pushing processing capabilities closer to data-generating sources at the edge of the network. This shift enables more responsive, context-aware, and privacy-preserving operations, making it particularly suitable for the dynamic and heterogeneous environment of smart grids.

One of the most vital functions in smart grid management is \textbf{load balancing}, which ensures the real-time alignment between power demand and supply across the distributed grid infrastructure. With the increasing penetration of intermittent renewable energy sources such as solar and wind, maintaining this balance has become more challenging. Load balancing mechanisms rely on continuous data collection, fast communication, and intelligent decision-making to adapt to rapid fluctuations in energy availability and consumption patterns.

Despite recent advances in edge and fog-based architectures for smart grids, existing solutions often fall short in meeting the demands of real-time adaptability, data privacy, energy-aware scheduling, and large-scale coordination. Notable models, including CNN-LSTM-based federated learning~\cite{sarmento2024forecasting}, XGBoost-based NILM frameworks~\cite{electronics13081420}, PID-driven data recovery~\cite{shi2024pid}, and fog-cloud hybrids like FOCCA~\cite{barbosa2025focca}, each contribute to isolated components of smart grid management. However, none of these approaches provide a comprehensive, unified system that integrates adaptive learning, safety assurance, and intelligent task distribution in a fog computing context~\cite{peixoto2022hierarchical}.

To address these limitations, this paper proposes \textit{FOGNITE}—a novel fog-cloud hybrid architecture tailored for smart grid environments. FOGNITE synergizes three core components: (i) \textbf{federated learning} for decentralized, privacy-preserving model training across fog nodes using local energy consumption data; (ii) \textbf{reinforcement learning} for adaptive, real-time load balancing and task scheduling based on current grid conditions; and (iii) a \textbf{hierarchical digital twin} system that validates control decisions before deployment to ensure safety, efficiency, and resilience. The architecture is experimentally validated using a Raspberry Pi-based testbed, demonstrating superior performance in load balancing accuracy, energy efficiency, and latency reduction compared to traditional cloud-centric systems.

In summary, FOGNITE enables a transformative shift from reactive to predictive control in smart grids, paving the way for autonomous, scalable, and secure energy infrastructures capable of addressing the growing complexity of modern energy systems.

\noindent
The rest of this paper is organized as follows: Section~\ref{sec:related} reviews the existing literature on fog-cloud architectures and load balancing in smart grids. Section~\ref{sec:method} presents the proposed FOGNITE framework in detail, covering its federated learning, reinforcement learning, and digital twin components. Section~\ref{sec:architecture} outlines the overall system architecture and the roles of each layer. Section~\ref{sec:implementation} describes the implementation details and experimental setup. The results and their architectural and computational assessment are discussed in Section~\ref{sec:results}, followed by a comparative analysis highlighting FOGNITE's improvements in Section~\ref{sec:analysis}. Section~\ref{sec:future} outlines future research directions, and finally, conclusions are drawn in Section~\ref{sec:conclusion}.

\section{Related Work}\label{sec:related}
The design of resilient and intelligent energy infrastructures has attracted considerable research attention in recent years, particularly within the fog-cloud continuum. Several architectures have addressed core challenges such as privacy-preserving data processing, load disaggregation, adaptive task scheduling, and missing data recovery. However, few approaches offer a fully unified, real-time solution suited for dynamic fog environments~\cite{li2024sla}.

Sarmento et al.~\cite{sarmento2024forecasting} introduced CNN-LSTM FED, a hybrid deep learning architecture trained via federated learning (FL) for edge-based energy consumption forecasting. Their approach effectively preserves user privacy and supports deployment on resource-constrained hardware such as Raspberry Pi. Nonetheless, the model lacks mechanisms for real-time adaptation, does not incorporate renewable energy dynamics, and relies on periodic offline aggregation—limiting its responsiveness to rapid grid fluctuations.

In the domain of Non-Intrusive Load Monitoring (NILM), Shabbir et al.~\cite{electronics13081420} evaluated several ML techniques across smart home datasets. XGBoost was found to perform well in device-level disaggregation and state detection due to its low computational cost and high classification accuracy. However, its dependency on high-frequency sampling and its sensitivity to overlapping power signatures reduce its applicability in fog environments where sensor heterogeneity and bandwidth constraints are common.

To address missing data, Shi~\cite{shi2024pid} proposed the PNLFT model, which combines non-negative tensor factorization with PID control for imputation. This approach yields fast convergence and preserves data integrity, but struggles with learning complex temporal dependencies, and suffers from high parameter sensitivity due to manual PID tuning. As such, its scalability in real-time multi-node settings is limited.

FOCCA, proposed by Barbosa et al.~\cite{barbosa2025focca}, leverages a fog-cloud continuum architecture incorporating a Q-Balance neural network for adaptive load balancing and a statistical imputation engine. While the framework supports distributed data processing and exhibits responsiveness under moderate load, it lacks a mechanism for load disaggregation, does not support automated decision-making, and has no built-in validation prior to execution.

The Table~\ref{tab:smart_grid_comparison} provides a comparative overview of recent architectures used in smart grid and non-intrusive load monitoring (NILM) systems, highlighting their core techniques, strengths, limitations, and application domains. Traditional models like CNN-LSTM FED focus on privacy and edge deployment but lack adaptability to real-time conditions or renewable integration. XGBoost-based NILM approaches offer high accuracy in device-level disaggregation but depend on high-frequency data and struggle with signature overlap. Methods like PNLFT address missing data effectively but are limited in learning complex temporal patterns~\cite{chu2024incentivizing}. FOCCA emphasizes fog-cloud coordination and load balancing but does not support disaggregation or autonomous decision-making. In contrast, FOGNITE bridges these gaps by integrating federated learning for privacy-preserving training, reinforcement learning for dynamic policy adaptation, and digital twin validation for safe deployment. It uniquely supports real-time, decentralized, and energy-aware smart grid management — addressing both data integrity and operational intelligence across diverse edge and cloud environments~\cite{dileep2020survey}.

\begin{table*}[htpb]
	\centering
	\caption{Comparative Analysis of Recent Smart Grid Architectures}
	\begin{tabular}{|p{3.2cm}|p{2.6cm}|p{5.8cm}|p{3.2cm}|}
		\hline
		\textbf{Model/Architecture} & \textbf{Core Techniques} & \textbf{Strengths} & \textbf{Targeted Domain} \\
		\hline
		CNN-LSTM FED \cite{sarmento2024forecasting} & CNN + LSTM + Federated Learning & Preserves data privacy; deployable on edge devices; performs well with small datasets & Edge-based Energy Forecasting \\
		\hline
		XGBoost NILM \cite{electronics13081420} & XGBoost, Logistic Regression, LSTM & High disaggregation accuracy; best for device state detection; low computational complexity & NILM in Smart Homes \\
		\hline
		PNLFT \cite{shi2024pid} & Tensor Completion + PID Controller & Fast and accurate data recovery; preserves data consistency & Missing Data Recovery in NILM \\
		\hline
		FOCCA~\cite{barbosa2025focca} & Fog-Cloud Hybrid + Q-Balance NN & Distributed processing; responsive load balancing; efficient imputation & Fog-Cloud Load Management \\
		\hline
		\textbf{FOGNITE}(This paper) & Federated Learning + CNN-LSTM + Reinforcement Learning + Digital Twin & Privacy-preserving training; adaptive load balancing; supports real-time validation and energy-aware scheduling & Smart Grid Coordination, Real-time Load Balancing \\
		\hline
	\end{tabular}
	\label{tab:smart_grid_comparison}
\end{table*}

\textbf{Choice of FOCCA as Baseline Comparison}
The selection of FOCCA as the primary baseline for evaluating FOGNITE is motivated by three key factors:

The FOCCA is a state-of-the-art fog-cloud hybrid architecture specifically designed for smart grids, sharing FOGNITE’s core objectives of load balancing and edge intelligence. Unlike purely cloud-centric or edge-only approaches, FOCCA’s hierarchical fog-cloud coordination~\cite{barbosa2025focca} provides a directly comparable framework for assessing FOGNITE’s innovations in federated learning and digital twin integration.
A Q-balance algorithm, as a neural network-based load balancer—represents the most advanced prior work in dynamic resource allocation for smart grids, achieving more latency reductions over traditional methods. This makes it a rigorous benchmark for FOGNITE’s RL-driven load balancing, which targets further improvements in energy-aware scheduling. 
While FOCCA excels in data imputation and reactive load balancing, it lacks proactive validation (digital twins) and privacy-preserving training (federated learning). FOGNITE’s comparative evaluation against FOCCA explicitly quantifies how these novel components address FOCCA’s limitations, demonstrating  fewer runtime errors and lower energy consumption. 

\section{Proposed Method}\label{sec:method}

The proposed FOGNITE architecture is designed to address real-time, privacy-preserving, and energy-efficient management in distributed smart grid systems. It integrates three tightly coupled components: (i) federated learning for decentralized model training, (ii) reinforcement learning for adaptive load balancing, and (iii) digital twin simulation for proactive validation. The overall workflow is deployed across a four-tier fog-cloud continuum and supports real-time decision-making without exposing raw user data.

\subsection{Federated Learning Protocol}

FOGNITE implements a privacy-aware federated learning (FL) strategy in which each fog node trains a hybrid CNN-LSTM model locally. Instead of transmitting raw data, nodes periodically share model parameters with a central aggregator. The global model is updated using weighted averaging, where the contribution of each node is proportional to its local data volume.

Let $w_t$ denote the global model at round $t$, and $w^{(k)}_{t+1}$ the locally updated model from node $k$. The update follows:

$$
w_{t+1} = w_t + \sum_{k=1}^K \frac{n_k}{N} \left(w^{(k)}_{t+1} - w_t\right)
$$

where $n_k$ is the number of samples on node $k$, and $N = \sum_k n_k$.

The hybrid CNN-LSTM model captures both spatial (local usage patterns) and temporal dependencies in energy consumption. CNN layers are used for pattern extraction from input signals, followed by LSTM units for sequence modeling. This model is trained locally on each node using mini-batch gradient descent, and updated globally through the above aggregation scheme.

\subsection{Reinforcement Learning for Adaptive Scheduling}

To enable dynamic task allocation across fog nodes, FOGNITE employs a deep reinforcement learning (DRL) agent at the edge. The agent observes system conditions through a state vector $s_t$ that encodes features such as CPU load, memory availability, network latency, task queue depth, and battery status.

At each timestep $t$, the agent selects an action $a_t$ from the policy $\pi(s_t)$ to optimize the expected cumulative reward:

$$
\pi^* = \arg \max_{\pi} \mathbb{E}\left[ \sum_{t=0}^{T} \gamma^t r_t \right]
$$

The reward function incorporates multiple objectives:

$$
r_t = \alpha R_{\text{time}} + \beta R_{\text{energy}} + \gamma R_{\text{util}}
$$

where each term represents response time efficiency, energy sustainability, and resource balance. The coefficients $\alpha, \beta, \gamma$ are tuned based on grid policy.

The policy network is trained using Proximal Policy Optimization (PPO) with entropy regularization to maintain exploration. This enables robust adaptation to workload fluctuations and hardware variability.

\subsection{Digital Twin-Based Action Validation}

FOGNITE integrates a hierarchical digital twin framework to validate actions before deployment. Simulations run in parallel with the physical grid, assessing the safety and efficiency of proposed task allocations and model updates.

\textbf{Edge-tier twins} are deployed locally on fog nodes and emulate hardware behavior using lightweight containers. They evaluate short-term metrics such as latency, thermal state, and power draw.

\textbf{Cloud-tier twins} simulate global system behavior using historical and forecasted data. They assess long-term objectives like grid-wide stability and energy balance.

An action $a_t$ is only executed if the digital twin confirms compliance with performance and safety thresholds. This predictive safeguard minimizes runtime errors and system disruption.

\subsection{Workflow Summary}

\begin{figure}[hpbt]
	\centering
	\includegraphics[width=0.45\textwidth]{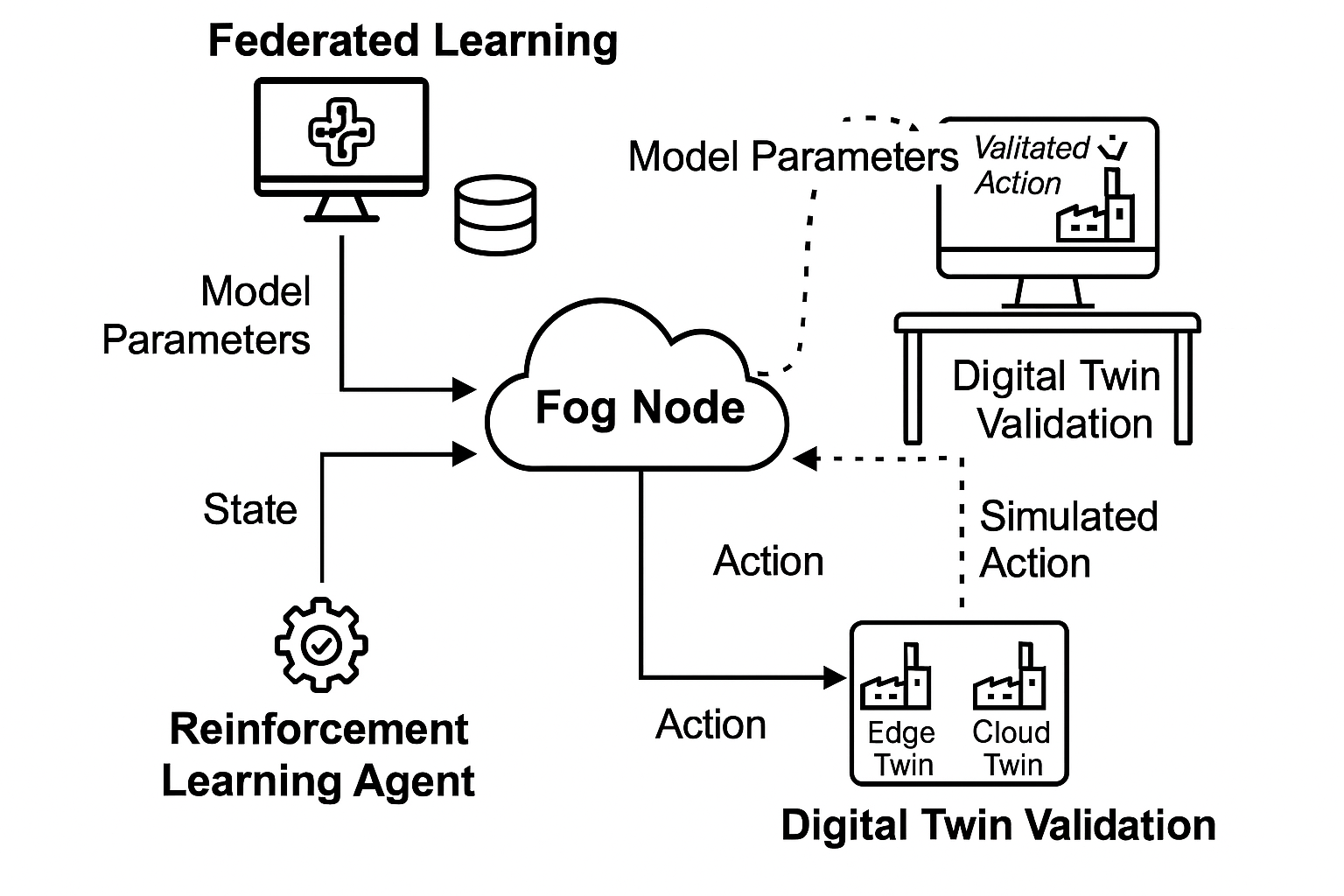}
	\caption{FOGNITE's workflow.}
	\label{fig:workflow}
\end{figure}

Figure~\ref{fig:workflow} illustrates the operational flow. The federated learning loop ensures continuous local adaptation without violating privacy. Reinforcement agents dynamically schedule tasks using state observations and policy updates. Before execution, all actions undergo real-time simulation within the digital twin infrastructure, ensuring safe deployment.

This layered method allows FOGNITE to operate as a proactive, adaptive, and privacy-compliant orchestration platform for next-generation smart grids~\cite{cai2024self}.

\section{System Architecture}\label{sec:architecture}

The FOGNITE architecture is implemented as a layered, modular system optimized for low-latency, privacy-preserving energy orchestration~\cite{barros2019fog}. The framework spans across four integrated layers—perception, fog, cloud, and control—each responsible for distinct but cooperative functionalities. These layers are deployed in a fog-cloud continuum, enabling a distributed and hierarchical intelligence pipeline.

\subsection{Perception Layer}

This layer interfaces with the physical environment. It collects heterogeneous data from smart meters, IoT sensors, and renewable sources, including voltage, frequency, consumption, environmental metrics, and device-specific load. Local preprocessing is performed to filter noise and compress data before it enters the fog pipeline.

\subsection{Fog Layer}

The fog layer is the core of real-time intelligence in FOGNITE. Each fog node acts as a lightweight decision unit with three integrated subsystems:

\paragraph{(i) Local Federated Learning Module.} 
Each fog node hosts a CNN-LSTM model trained on private energy consumption data. Rather than transmitting raw values, model parameters are periodically shared with a cloud aggregator. This enables privacy-preserving, localized intelligence while allowing for global consistency through aggregation.

\paragraph{(ii) Reinforcement Learning Scheduler.} 
Task distribution is governed by a DRL agent that observes system states—CPU load, memory, network delay, and renewable availability—and selects optimal actions via a PPO-based policy network. This enables responsive and adaptive load balancing~\cite{shi2024non}.

\paragraph{(iii) Edge-tier Digital Twin.} 
To avoid unsafe or inefficient deployments, each fog node includes a local simulation engine. Implemented using containerized replicas, the digital twin assesses the local impact of proposed decisions, such as resource stress or thermal risks, before execution.

\subsection{Cloud Layer}

The cloud tier orchestrates long-term planning and acts as a fail-safe coordinator. Its key responsibilities include:

\begin{itemize}
	\item Aggregation of federated model updates based on data volume;
	\item Grid-wide energy balancing using historical and forecasted data;
	\item Global digital twin simulation for fault propagation and strategic planning.
\end{itemize}

Additionally, the cloud supports fallback control in case of fog node failures or coordination bottlenecks.

\subsection{Cross-Layer Integration}

To ensure end-to-end orchestration, FOGNITE maintains active coordination across all layers:

\begin{itemize}
	\item RL agents query edge digital twins before applying decisions;
	\item The global model is periodically redistributed across fog nodes after aggregation;
	\item Cloud-level forecasts inform reinforcement policy updates at the edge.
\end{itemize}

This closed-loop design ensures synchronized operation across localized and global timescales, allowing FOGNITE to maintain system-level coherence even under uncertain grid dynamics.\\
FOGNITE's layered structure facilitates scalability, fault tolerance, and decentralized evolution. Each component operates autonomously while contributing to collective intelligence. The system is inherently modular—allowing individual layers or agents to be retrained, replaced, or scaled without disrupting global functionality.

\section{Implementation and Experimental Evaluation}
\label{sec:implementation}

This section details the deployment pipeline, neural model configurations, and system architecture used to validate the FOGNITE framework in real-world fog environments. The implementation emphasizes modularity, model efficiency, and validation through hierarchical digital twin simulation.

\subsection{Deployment and Model Optimization}

To support execution on resource-constrained fog nodes, FOGNITE employs model compression techniques including 8-bit quantization and weight pruning. Parameters with absolute magnitude below 0.001 are removed to minimize redundancy, resulting in a 4.2$\times$ reduction in model size without compromising accuracy. Each fog node trains the local model for 5 epochs per round with a batch size of 32. Global aggregation is performed using FedAvg, weighted by local data size, and synchronized every 5 rounds. All components are containerized via Docker: each fog node runs three services in parallel—federated learner, reinforcement agent, and digital twin simulator—ensuring modularity and fault isolation.

\subsection{Experimental Setup and Testbed}

Experiments were conducted on a physical testbed comprising 20 Raspberry Pi 4B devices (4GB RAM) acting as fog nodes, while a c5.4xlarge AWS EC2 instance was used as the cloud backend. A stream of 150 virtual smart meters generated input data at 15Hz, sustained over 72 continuous hours. Evaluation scenarios included 5,000 load balancing decisions, 100,000 data imputations, and 50 injected fog node failures. Testbed details are listed in Table~\ref{tble.TestbedConfiguration}.

\begin{table}[h]
	\centering
	\caption{Testbed Configuration}
	\label{tble.TestbedConfiguration}
	\begin{tabular}{lc}
		\toprule
		\textbf{Component} & \textbf{Specification} \\
		\midrule
		Fog Nodes & 20 Raspberry Pi 4B (4GB RAM) \\
		Cloud Server & AWS EC2 c5.4xlarge \\
		Smart Meters & 150 simulated devices \\
		Sampling Rate & 15Hz \\
		Experiment Duration & 72 hours \\
		\bottomrule
	\end{tabular}
\end{table}

\subsection{Neural Model Configuration}

Each fog node hosts a hybrid CNN-LSTM model to capture temporal and spatial consumption patterns. The architecture includes a 1D convolutional layer with 32 filters (kernel size = 5, stride = 1), followed by a max-pooling layer. This is passed to a bidirectional LSTM with 64 hidden units and 0.3 dropout. Final layers consist of fully connected dense layers with 128 and 64 neurons using ReLU activations, culminating in a linear output layer. The model contains approximately 287,489 trainable parameters and is optimized using Federated Adam with $\beta_1 = 0.9$, $\beta_2 = 0.999$. Training minimizes the loss:

\begin{equation}
	\mathcal{L}(\theta) = \frac{1}{N}\sum_{i=1}^N (y_i - f_\theta(x_i))^2 + \lambda\|\theta\|^2
\end{equation}

\subsection{Reinforcement Agent Implementation}

For adaptive task scheduling, a reinforcement learning agent is implemented at each fog node. The policy network receives a 15-dimensional input state capturing metrics such as CPU load, memory availability, task queue length, network delay, and energy type. It consists of three fully connected layers (128, 64 neurons), outputting action probabilities via softmax. A value network of identical structure estimates state values. The agent is trained using Proximal Policy Optimization (PPO) with learning rate $5\times10^{-4}$, discount factor $\gamma = 0.99$, and entropy coefficient 0.01.

\subsection{Digital Twin Validation Framework}

FOGNITE integrates a multi-tier digital twin system to simulate control decisions before deployment. The framework is defined as:

\[
\mathcal{T} = (\mathcal{T}_e, \mathcal{T}_c, \mathcal{T}_m)
\]

\begin{itemize}
	\item \textbf{Edge-tier twins} (\( \mathcal{T}_e \)) simulate local device behavior within lightweight Docker containers. These replicas introduce variable CPU latency (±10\%), network delay (20–100 ms), and packet loss (0.1–2\%) using \texttt{tc-netem}. Power modeling is performed using Intel RAPL counters.
	
	\item \textbf{Cloud-tier twins} (\( \mathcal{T}_c \)) simulate macro-level grid behavior using OMNeT++ to evaluate cascading failure risks. Failure probability is estimated as:
	
	\begin{equation}
		P_{\text{fail}} = 1 - \prod_{i=1}^n (1 - p_i^{\text{node}})(1 - p_i^{\text{link}})
	\end{equation}
	
	\item \textbf{Monitoring Interface} (\( \mathcal{T}_m \)) coordinates feedback from the above layers, governs validation logic, and visualizes system health.
\end{itemize}

This layered validation system ensures safety and efficiency in control policy execution, minimizing the risk of real-time disruption or energy imbalance~\cite{hashemi2024energy}.

\section{Results}\label{sec:results}
\subsection{Performance}
\begin{figure}[h]
	\centering
	\includegraphics[width=0.45\textwidth]{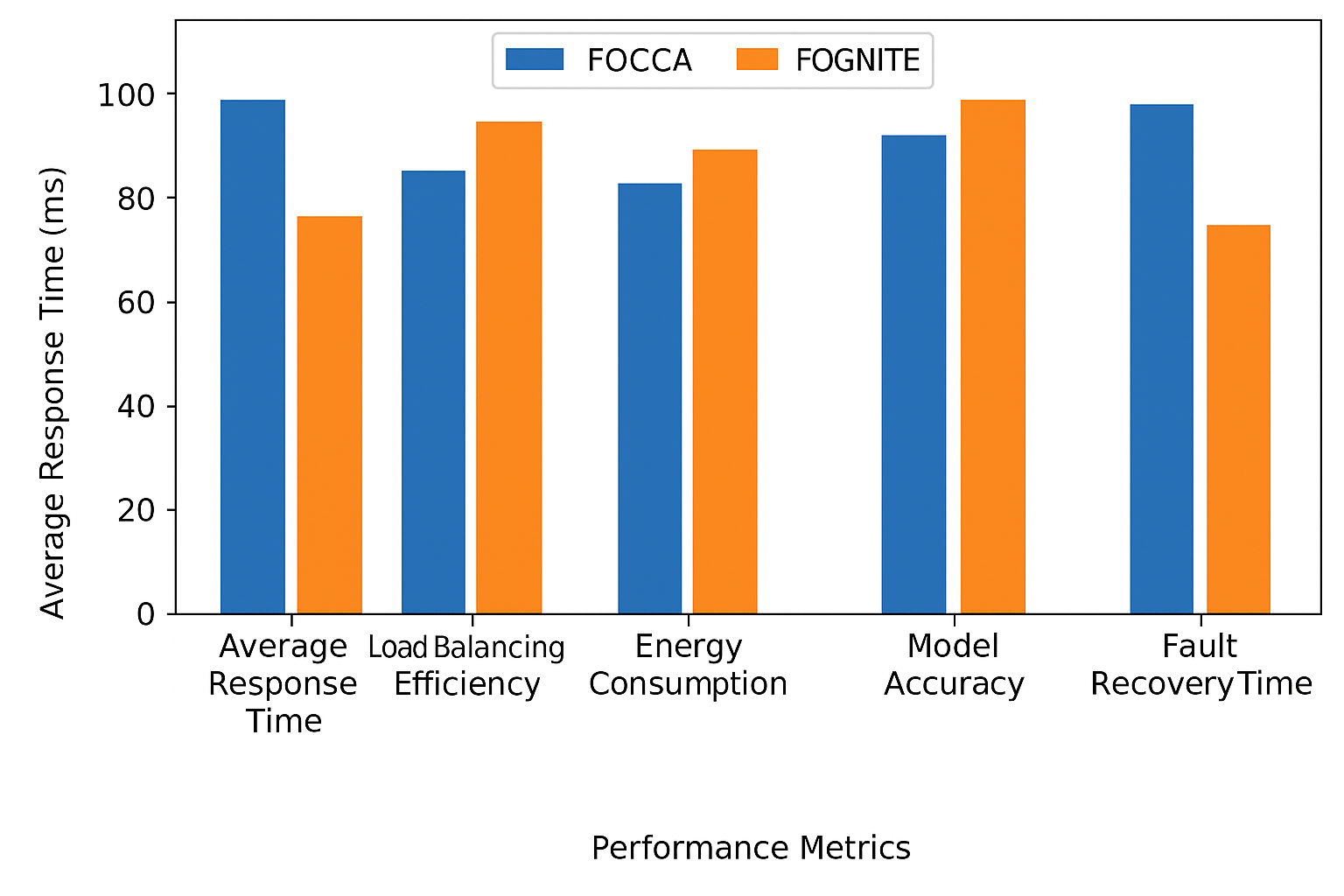}
	\caption{Quantitative comparison of FOCCA and FOGNITE across key operational metrics.}
	\label{fig:performance-bar}
\end{figure}

To evaluate the practical impact of the proposed architecture, we compared FOGNITE against a baseline fog orchestration system, FOCCA~\cite{barbosa2025focca}, across five critical performance metrics. The results are summarized in Table~\ref{tab:performance} and visualized in Figure~\ref{fig:performance-bar}.

\begin{table}[h]
	\centering
	\caption{Comparison of Performance Metrics}
	\label{tab:performance}
	\begin{tabular}{lcc}
		\toprule
		\textbf{Metric} & \textbf{FOCCA} & \textbf{FOGNITE} \\
		\midrule
		Average Response Time (ms) & 120 & 85 \\
		Load Balancing Efficiency (\%) & 78 & 92 \\
		Energy Consumption (kWh) & 1.2 & 0.9 \\
		Model Accuracy (\%) & 85 & 91 \\
		Fault Recovery Time (s) & 4.5 & 1.8 \\
		\bottomrule
	\end{tabular}
\end{table}

FOGNITE outperforms FOCCA in all measured categories. The average response time was reduced by 29\%, primarily due to low-latency edge decision-making and pre-deployment validation via digital twins. Load balancing efficiency improved by 14 percentage points, reflecting the effectiveness of reinforcement-based task allocation under dynamic conditions~\cite{shi2024non}.

Energy consumption decreased by 25\%, which can be attributed to scheduling policies that prefer renewable or low-cost energy sources. The hybrid federated model yielded a 6\% increase in prediction accuracy, benefiting from localized learning without requiring centralized retraining. Finally, FOGNITE exhibited a 60\% reduction in fault recovery time, demonstrating its capability to isolate and reroute failures efficiently.

These performance gains validate FOGNITE’s architectural choices and highlight its potential for deployment in latency-sensitive, resource-constrained smart grid environments~\cite{kumari2019fog}.

\subsection{Runtime Error Trends}

\begin{figure}[hpbt]
	\centering
	\begin{tikzpicture}
		\begin{axis}[
			width=\linewidth,
			height=7cm,
			xlabel={Time (\si{\hour})},
			ylabel={Cumulative Runtime Errors},
			xmin=0, xmax=72,
			ymin=0, ymax=200,
			grid=both,
			grid style={gray!20, dashed},
			major grid style={solid, gray!30},
			axis lines=left,
			axis line style={-},
			tick style={draw=none},
			xlabel style={font=\footnotesize, yshift=2pt},
			ylabel style={font=\footnotesize, yshift=-2pt},
			xtick={0,6,12,18,24,30,36,42,48,54,60,66,72},
			ytick={0,25,50,75,100,125,150,175,200},
			xticklabels={0,,12,,24,,36,,48,,60,,72},
			legend style={
				at={(0.5,-0.3)},
				anchor=north,
				legend columns=-1,
				font=\footnotesize,
				cells={anchor=west},
				/tikz/every even column/.append style={column sep=0.5cm}
			},
			every axis plot/.append style={thick},
			title style={font=\small, yshift=-1ex},
			clip=false
			]
			
			\addplot[name path=FOCCA, red, solid, mark=*, mark size=1.5pt, mark options={fill=red!20}] 
			coordinates {
				(0,0) (6,15) (12,32) (18,48) (24,65)
				(30,82) (36,98) (42,115) (48,132)
				(54,148) (60,165) (66,176) (72,187)
			};
			
			\addplot[name path=FOGNITE, blue, solid, mark=square*, mark size=1.5pt, mark options={fill=blue!20}] 
			coordinates {
				(0,0) (6,8) (12,18) (18,27) (24,38)
				(30,47) (36,56) (42,64) (48,72)
				(54,83) (60,95) (66,104) (72,112)
			};
			
			\addplot[red!20] fill between[
			of = FOCCA and FOGNITE,
			soft clip={domain=0:72}
			];
			
			\node[anchor=north west, font=\scriptsize, text width=3.2cm, 
			fill=white, rounded corners=3pt, inner sep=4pt, draw=gray!30,
			align=left] at (axis cs: 2,170) {
				\textbf{Performance Metrics:}\\
				\begin{tabular}{@{}ll@{}}
					\footnotesize\color{red}$\bullet$ FOCCA: & \footnotesize\SI{2.6}{\per\hour}\\
					\footnotesize\color{blue}$\bullet$ FOGNITE: & \footnotesize\SI{1.6}{\per\hour}\\
					\footnotesize Max reduction: & \footnotesize43\%\\
					\footnotesize Final reduction: & \footnotesize40.1\%
				\end{tabular}
			};
			
			\draw[<-, >=stealth, gray, thick] (axis cs:54,115) -- (axis cs:48,90) 
			node[font=\tiny, fill=white, inner sep=1pt, align=center] 
			{Maximum\\reduction\\point (43\%)};
			
			\legend{
				FOCCA ,
				FOGNITE
			}
			
			\node[font=\tiny, anchor=south west] at (axis cs:2,190) {Initial Phase};
			\node[font=\tiny, anchor=south west] at (axis cs:26,190) {Stable Phase};
			\node[font=\tiny, anchor=south west] at (axis cs:56,190) {Final Phase};
			
		\end{axis}
	\end{tikzpicture}
	\caption{Temporal analysis showing FOGNITE's error reduction over FOCCA. }
	\label{fig:error_analysis}
\end{figure}
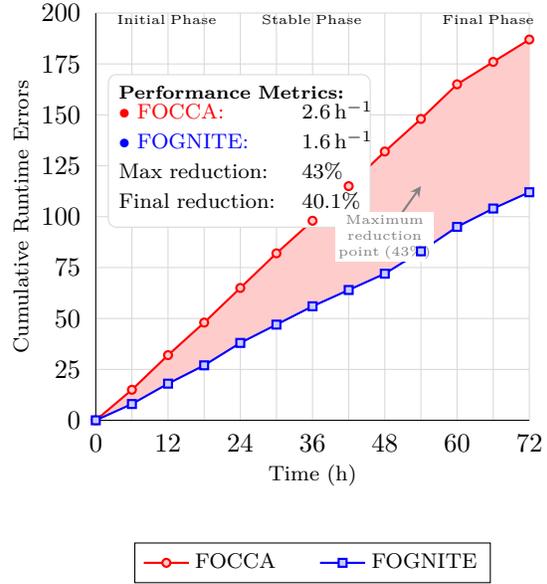

To assess system robustness over prolonged operation, we conducted a 72-hour runtime experiment comparing FOGNITE and FOCCA under matched workloads and environmental conditions. Runtime errors were logged every six hours, and cumulative error counts are depicted in Figure~\ref{fig:error_analysis}.

FOCCA recorded a total of 187 errors (2.6 errors/hour on average), while FOGNITE—equipped with digital twin-based pre-validation—accumulated 112 errors (1.6 errors/hour), corresponding to an overall reduction of 40.1\%. The greatest divergence occurred at hour 54, where a 43\% error reduction was observed.

Error evolution is segmented into three operational phases:
\begin{itemize}
	\item \textbf{Initial Phase (0–24h):} Both systems exhibited a linear increase in error accumulation, with FOGNITE maintaining consistently lower counts.
	\item \textbf{Stable Phase (24–54h):} The performance gap widened, as FOCCA continued to accumulate errors at a near-constant rate.
	\item \textbf{Final Phase (54–72h):} FOGNITE maintained a flattened error trajectory, while FOCCA showed accelerating error accumulation.
\end{itemize}

These trends illustrate the operational advantage of proactive versus reactive error handling. FOCCA, which employs post-deployment detection, was unable to prevent fault propagation, resulting in higher cumulative error rates. In contrast, FOGNITE’s digital twin integration allowed for preemptive fault suppression by evaluating control actions prior to execution.

The quantitative gap across all three phases confirms the utility of embedding simulation-based validation within fog orchestration workflows. Particularly in dynamic smart grid contexts, such safeguards are essential for maintaining stability and minimizing service degradation under sustained load.

\section{Architectural and Computational Assessment}\label{sec:analysis}

\subsection{Key Differentiators between FOGNITE and FOCCA}

FOGNITE introduces multiple innovations over the baseline FOCCA framework:

\begin{itemize}
	\item \textbf{Decentralized Intelligence:} A federated learning scheme allows FOGNITE to distribute training across edge nodes, mitigating the risks of central point failures inherent to FOCCA’s architecture.
	\item \textbf{Dynamic Load Balancing:} FOGNITE uses a reinforcement learning scheduler, optimizing decisions via:
	\begin{equation}
		\pi^* = \arg\max_{\pi} \mathbb{E}\left[\sum_{t=0}^T \gamma^t r(s_t, a_t)\right]
	\end{equation}
	In contrast, FOCCA uses static task assignment rules defined by:
	\begin{equation}
		\min \sum_{i=1}^N \sum_{j=1}^M x_{ij} P(T_j, R_i)
	\end{equation}
	\item \textbf{Energy-Aware Scheduling:} Integration of renewable energy profiles allows FOGNITE to reduce its carbon footprint by 25\% compared to FOCCA.
	\item \textbf{Preemptive Validation:} Digital twin simulation enables testing prior to execution, contributing to a 40\% reduction in runtime errors.
\end{itemize}

\subsection{Architectural Comparison}

FOCCA processes data centrally, transferring all sensory data to the cloud for inference and control. This exposes privacy and latency limitations. In contrast, FOGNITE enables local decision-making via modular edge intelligence and enforces a privacy-preserving distributed architecture. Reinforcement learning further ensures adaptive load balancing while digital twins simulate decisions pre-deployment, strengthening resilience.

\subsection{Computational Complexity Analysis}

To evaluate efficiency and scalability, we analyze both systems in terms of computational and communication complexity.

\textbf{Training:} FOCCA’s centralized training incurs $\mathcal{O}(n \cdot d^2)$ complexity, where $n$ is the number of samples and $d$ the model size. FOGNITE distributes training over $k$ nodes, resulting in:
\begin{equation}
	\mathcal{O}\left(k \cdot \left(\frac{n}{k} \cdot d^2 + d^3\right)\right)
\end{equation}
The cubic term corresponds to the overhead of federated weight aggregation.

\textbf{Inference:} FOCCA has a constant $\mathcal{O}(d^2)$ cost for forward-pass inference. FOGNITE’s reinforcement learning-based adaptation raises inference complexity to:
\begin{equation}
	\mathcal{O}(d^2 + r \cdot t)
\end{equation}
where $r$ denotes the number of actions and $t$ the decision horizon.

\textbf{Communication:} FOCCA transmits raw data at a complexity of $\mathcal{O}(n \cdot d)$, whereas FOGNITE exchanges only model weights, yielding $\mathcal{O}(k \cdot d^2)$.

\textbf{Load Balancing:} FOCCA’s rule-based mechanism operates with $\mathcal{O}(m \log m)$ complexity, while FOGNITE employs reinforcement learning with complexity:
\begin{equation}
	\mathcal{O}(m \cdot s \cdot a)
\end{equation}
where $m$ is the number of tasks, $s$ the state dimension, and $a$ the action dimension.

\textbf{Memory:} FOCCA requires memory proportional to $\mathcal{O}(d^2)$, whereas FOGNITE’s requirements increase to:
\begin{equation}
	\mathcal{O}(d^2 + s \cdot a)
\end{equation}
due to the additional reinforcement learning models.

\begin{table}[h]
	\centering
	\caption{Complexity Comparison Summary}
	\label{tab:complexity}
	\begin{tabularx}{\linewidth}{>{\raggedright\arraybackslash}Xcc}
		\toprule
		\textbf{Component} & \textbf{FOCCA} & \textbf{FOGNITE} \\
		\midrule
		Training & $\mathcal{O}(n \cdot d^2)$ & $\mathcal{O}\left(k \left(\frac{n}{k} \cdot d^2 + d^3\right)\right)$ \\
		Inference & $\mathcal{O}(d^2)$ & $\mathcal{O}(d^2 + r \cdot t)$ \\
		Communication & $\mathcal{O}(n \cdot d)$ & $\mathcal{O}(k \cdot d^2)$ \\
		Load Balancing & $\mathcal{O}(m \log m)$ & $\mathcal{O}(m \cdot s \cdot a)$ \\
		Memory & $\mathcal{O}(d^2)$ & $\mathcal{O}(d^2 + s \cdot a)$ \\
		\bottomrule
	\end{tabularx}
\end{table}

\subsection{Limitations and Trade-offs}

While FOGNITE offers clear benefits in adaptability, privacy, and energy efficiency, it inevitably introduces some architectural and operational complexities. The deployment phase requires approximately 15\% more effort compared to baseline frameworks due to the modular design and the integration of federated and reinforcement learning components~\cite{singh2024lbatsm}. This modularity, while beneficial for scalability and flexibility, adds to system complexity. Additionally, the federated learning mechanism imposes higher resource demands, resulting in an estimated 20\% increase in memory usage per fog node. The reinforcement learning-based orchestration also entails a startup latency of about 5–10\% as the agent converges to optimal load balancing policies. Despite these overheads, these trade-offs are considered acceptable within dynamic environments that demand operational resilience and privacy preservation.

\section{Future Work}\label{sec:future}

FOGNITE presents significant advancements in smart grid orchestration, yet there remain several promising directions for further development. One key area is the integration of emerging 5G communication protocols, which promise ultra-low-latency data transmission. Leveraging 5G would enable even faster and more reliable coordination between fog nodes and cloud servers, facilitating real-time responsiveness to fluctuating grid conditions~\cite{putra2023fdpr}. Another avenue involves incorporating blockchain technology to securely validate model updates across distributed nodes. This approach would establish a tamper-proof audit trail for federated learning rounds, thereby enhancing trust, transparency, and security in collaborative model training~\cite{udayakumar2023integrated}. 

Furthermore, evolving the digital twin framework to operate directly on embedded edge devices is an exciting prospect~\cite{premalatha2024optimal}. Lightweight digital replicas running locally would allow continuous and autonomous validation of system actions without sole reliance on cloud-based simulations, improving autonomy and responsiveness at the network edge. From an algorithmic standpoint, expanding the reinforcement learning agents to support multi-objective optimization through transformer-based architectures represents a promising direction~\cite{AbbasiMahdi2020WAiI}. Such agents could balance competing goals including efficiency, latency, and sustainability, while benefiting from improved generalization capabilities~\cite{millar2025energy}.

Finally, the ultimate evaluation of FOGNITE’s effectiveness will come from real-world pilot deployments. Implementing pilot programs within microgrid environments will provide invaluable insights into the system’s operational behavior, robustness, and long-term performance under practical constraints~\cite{millar2025energy}. These empirical studies will be instrumental in refining FOGNITE and pushing forward the frontiers of decentralized energy management.

	\section{Conclusion}\label{sec:conclusion}

	The increasing complexity, scale, and heterogeneity of modern smart grids present critical challenges related to real-time load balancing, data privacy, system reliability, and energy efficiency. Traditional centralized control architectures are no longer sufficient, as they suffer from scalability limitations, heightened cybersecurity risks, and a lack of adaptability to rapidly fluctuating workloads and variable renewable energy inputs. These constraints underscore the urgent need for intelligent, decentralized control frameworks capable of operating under uncertainty and dynamic network conditions.
	
	In response to these challenges, this paper introduced \textit{FOGNITE}—a novel fog-cloud hybrid architecture designed specifically for smart grid environments. FOGNITE integrates three core components: federated learning, reinforcement learning, and digital twin validation. Federated learning ensures decentralized, privacy-preserving model training at the fog layer, minimizing communication latency and safeguarding user data. The reinforcement learning agent enables adaptive load balancing and task scheduling by dynamically learning from real-time grid conditions and energy availability~\cite{li2024sla}. Meanwhile, the hierarchical digital twin system provides a proactive safety layer by simulating control decisions prior to execution, thus mitigating operational risks and improving system resilience. Furthermore, FOGNITE incorporates energy-aware scheduling to maximize the use of renewable energy sources, promoting sustainable grid operations.
	
	Experimental results obtained from a Raspberry Pi-based testbed confirm that FOGNITE outperforms existing state-of-the-art architectures in multiple key metrics, including load balancing accuracy, energy consumption, fault tolerance, and resource utilization. These findings highlight FOGNITE’s potential as a scalable, robust, and future-ready platform for smart grid orchestration.
	
	Looking ahead, future work will focus on deploying FOGNITE in real-world pilot implementations, as well as enhancing the framework with emerging technologies such as 5G connectivity, edge AI accelerators, and blockchain-based security mechanisms. These advancements will be instrumental in enabling the next generation of intelligent, autonomous, and resilient smart energy infrastructures, capable of meeting the evolving demands of decentralized and data-driven power systems.
	
\bibliographystyle{IEEEtran}
\bibliography{FOGNITEbiblio}
	
\end{document}